\documentclass[format=acmsmall]{acmart}

\usepackage{enumerate}
\usepackage[shortlabels]{enumitem}
\usepackage{graphicx}
\usepackage{subcaption}
\usepackage{amsmath,nccmath}
\usepackage{acronym}
\usepackage{booktabs}
\usepackage{xspace}
\usepackage{color}
\usepackage[ruled,linesnumbered,lined,boxed]{algorithm2e}
\usepackage{multirow}
\usepackage{lineno}

\AtBeginDocument{%
  \providecommand\BibTeX{{%
    \normalfont B\kern-0.5em{\scshape i\kern-0.25em b}\kern-0.8em\TeX}}}

\def\tsc#1{\csdef{#1}{\textsc{\lowercase{#1}}\xspace}}
\tsc{WGM}
\tsc{QE}
\tsc{EP}
\tsc{PMS}
\tsc{BEC}
\tsc{DE}

\newcommand{\ie}{i.e.,\xspace}
\newcommand{\eg}{e.g.,\xspace}

\newcommand{\RQ}[2]{
    \begin{description}[topsep=0pt,nosep=0pt, leftmargin=0.75cm, noitemsep,nolistsep]
    \phantomsection\label{section:setup:rq#1}
    \item[RQ#1] #2
    \end{description}
}
\newcommand{\RQRef}[1]{\textbf{\hyperref[section:setup:rq#1]{RQ#1}}}

\acrodef{TSCR}[TSCR]{Transformer-based Sequential Conversational Recommender}
\acrodef{CRS}[CRS]{Conversational Recommender System}
\acrodef{KG}[KG]{Knowledge Graph}
\acmJournal{TOIS}
\acmYear{2024}
\setcopyright{acmlicensed}
\copyrightyear{2024}
\acmDOI{XXXXXXX.XXXXXXX}

\begin{document}

\title{Knowledge-Enhanced Conversational Recommendation via Transformer-based Sequential Modelling}
                   
\author{Jie Zou}
\affiliation{%
 \institution{University of Electronic Science and Technology of China}
 \country{China}
 }
\email{jie.zou@uestc.edu.cn}

\author{Aixin Sun}
\affiliation{%
 \institution{SCALE@Nanyang Technological University}
 \country{Singapore}
 }
\email{AXSun@ntu.edu.sg}

\author{Cheng Long}
\affiliation{%
 \institution{SCALE@Nanyang Technological University}
 \country{Singapore}
 }
\email{c.long@ntu.edu.sg}

\author{Evangelos Kanoulas}
\affiliation{%
  \institution{University of Amsterdam}
  \country{The Netherlands}
}
\email{e.kanoulas@uva.nl}

\authorsaddresses{Jie Zou, Center for Future Media, School of Computer Science and Engineering, University of Electronic Science and Technology of China, China; 
jie.zou@uestc.edu.cn; Aixin Sun, SCALE@Nanyang Technological University, 50 Nanyang Ave, Singapore; Cheng Long, SCALE@Nanyang Technological University, 50 Nanyang Ave, Singapore; Evangelos Kanoulas, University of Amsterdam, Science Park 904, 1098 XH Amsterdam, The Netherlands, e.kanoulas@uva.nl.}
\thanks{This work was partly done when Jie Zou was a research fellow at SCALE@Nanyang Technological University, Singapore.}

\renewcommand{\shortauthors}{Jie Zou, et al.}

\begin{abstract}
In \acp{CRS}, conversations usually involve a set of items and item-related entities or attributes, \eg director is a related entity of a movie. These items and item-related entities are often mentioned along the development of a dialog, leading to potential sequential dependencies among them. However, most of existing \acp{CRS} neglect these potential sequential dependencies. In this paper, we first propose a Transformer-based sequential conversational recommendation method, named TSCR, to model the sequential dependencies in the conversations to improve \ac{CRS}. In TSCR, we represent conversations by items and the item-related entities, and construct user sequences to discover user preferences by considering both the mentioned items and item-related entities. Based on the constructed sequences, we deploy a Cloze task to predict the recommended items along a sequence. Meanwhile, in certain domains, knowledge graphs formed by the items and their related entities are readily available, which provide various different kinds of associations among them. Given that TSCR does not benefit from such knowledge graphs, we then propose a knowledge graph enhanced version of TSCR, called TSCRKG. In specific, we leverage the knowledge graph to offline initialize our model TSCRKG, and augment the user sequence of conversations (\ie sequence of the mentioned items and item-related entities in the conversation) with multi-hop paths in the knowledge graph. Experimental results demonstrate that our TSCR model significantly outperforms state-of-the-art baselines, and the enhanced version TSCRKG further improves recommendation performance on top of TSCR.
\end{abstract}



\maketitle

\section{Introduction}
\label{sec:intro}
In general, a recommender system learns user preference from historical user-item interactions and then recommends items based on the learned user preference. The recommended items can be delivered to users through various interfaces, \eg a list of products on e-commerce websites. Thanks to the rapid development of chatbots, \ac{CRS} is now becoming a promising interface to deliver recommended items to users directly through dialogs. %

An example dialog for \ac{CRS} is shown in Fig.~\ref{figure:dialogExample}. The task of this example dialog is to recommend movies (\ie the items) to users. Observe that there are two important properties demonstrated in this dialog: (a) first, it is common to mention entities that are closely related to the items to be recommended, \eg director and genre of a movie in our example; (b) second, the mentioned items (and also item-related entities) naturally form a sequence, following the development of the conversation. 
That is, there is an order impact among items in a conversation and potential sequential dependencies exist within the conversation. For instance, when people talk about a movie (\eg Fast \& Furious 1), it is natural and reasonable to recommend a sequel or prequel of that movie (\eg Fast \& Furious 4). Motivated by this observation, we argue that the modeling of such sequential dependency can well capture the context of users' activities,  and has great potential to improve the quality of recommendations. 

Generally, a~\ac{CRS} integrates two modules: a recommender module, and a dialog module. 
The latter generates natural language conversations to interact with users. The recommender module focuses on recommending desirable items to users by utilizing the information from the conversation, as well as related information from external sources like historical user-item interactions and knowledge bases. In this work, we focus on the recommender module only.

\begin{figure}[t]
\centering
 \small
\begin{tabular}{l|p{2.3in}}
\toprule
\textbf{Role} & \textbf{Message}\\
\midrule 
Seeker & Hi I am looking for a good \textcolor{red}{thriller} really, any type new or old.\\
Recommender & Hi there! Do you want \textcolor{red}{action} or \textcolor{red}{suspense}?\\
Seeker & Yes I do. \\
Recommender & My favorite \textcolor{red}{thrillers} are the sequels to \textcolor{blue}{Fast \& Furious 1}.\\
Seeker & Yeah that one is good.\\
Recommender & The best one is the fourth one. It's called \textcolor{blue}{Fast \& Furious 4}, it's full of \textcolor{red}{action} with an addicting plot.\\
Seeker & I love those. Have you seen \textcolor{blue}{Fast \& Furious 8} yet?\\
\dots & \dots\\
\bottomrule
\end{tabular}
\caption{An example dialog from the ReDial dataset. The mentioned items (\ie movies) are highlighted in blue color, and item-related entities in red color.}
\label{figure:dialogExample}
\end{figure}

Conversational systems have shown great potential in a wide range of areas, such as information-seeking systems \citep{radlinski2017theoretical}, conversational product search \citep{zou2022learning}, dialog systems \citep{stoyanchev2014towards, de2003analysis}, and \ac{CRS} \citep{SepliarskaiaKRR18, zou2022improving-sigir}. As for \ac{CRS}, a number of solutions have been proposed~\citep{ren2022variational, manzoor2021generation, jannach2020end, zou2020towardsb}, given it is a rapidly growing research topic in recent years. Early efforts in the development of CRS can be traced back to interactive recommendations ~\cite{he2016interactive, jugovac2017interacting} and question-based recommendations~\cite{zou2020towardsb}. More recently, two categories of \acp{CRS} including anchor based \acp{CRS} \citep{zou2020towardsb} and dialog based \acp{CRS} \citep{zou2022improving-sigir} have been proposed. Anchor based \acp{CRS} is based on the ``system asks -- user responds'' mode and simulates conversations by using some ``anchor'' text, \eg item aspects~\citep{zhang2018towards}, facets~\citep{sun2018conversational},  entities~\citep{zou2020towardsb, zou2019learning, zou2020towardsa}, topics \citep{christakopoulou2018q}, and attributes~\citep{lei2020estimation,lei2020interactive,zhou2020leveraging}. They usually utilize a belief tracker to infer the anchor-based preferences to improve \ac{CRS}. Another mainstream approach, dialog based \acp{CRS}, is based on human-generated dialogs~\citep{chen2019towards,zhou2020improving,sarkar2020suggest,CRWalker,zhou2020towards,wang2021conversation,wang2022recindial,chen2021knowledge,yang2021improving,zhang2021kecrs,lu2021revcore}. For instance, at the early stage, \citet{li2018towards} proposed a \ac{CRS} dataset, ReDial, and presented a benchmark model for item recommendation. ReDial soon became the most widely used dataset for \ac{CRS}. Subsequently, the dataset TG-ReDial is proposed for \ac{CRS} in the Chinese scenario, which is constructed in a similar way with ReDial \citep{zhou2020towards}. Given that entities within utterances of ReDial and TG-ReDial are linked to a knowledge base, most subsequent work uses knowledge graphs to improve \ac{CRS}~\citep{chen2019towards,zhou2020improving,sarkar2020suggest,CRWalker,zhou2020towards, zhou2021crfr}. 
Although the aforementioned work demonstrates success to some extent, they neglect the \textit{order impact} among items or entities discussed earlier. Hence, these existing solutions do not model the potential sequential dependency within the conversations. 

Inspired by the success of Transformer methods like BERT~\citep{devlin2018bert}, in our research, we explicitly model the bidirectional sequential dependency in conversations by using Transformer. Recent studies have shown that a carefully designed task-specific input format to BERT could lead to state-of-the-art performance~\citep{xue2022embarrassingly}. Our solution is along this line. 
We first propose a simple and effective \acf{TSCR} model for \acp{CRS}~\citep{zou2022improving-sigir}. Specifically, \ac{TSCR} extracts both the mentioned items and item-related entities in conversations as contextual information to construct a user sequence. 
Based on this user sequence, \ac{TSCR} randomly masks some items and deploys a Cloze task~\citep{devlin2018bert, taylor1953cloze} to predict the masked items by leveraging the bidirectional contextual information in the input sequence. The bidirectional representation for the user sequence is modeled by the deep bidirectional self-attention architecture. 

Given that the use of knowledge graphs has shown to be highly beneficial for recommender systems~\citep{zhou2020improving,sarkar2020suggest,CRWalker, zhou2021crfr,deng2021unified},
we further propose a Transformer-based Sequential Conversational Recommender enhanced by knowledge graphs (TSCRKG), as an extension of \ac{TSCR}. Knowledge graphs provide additional information such as closely related items/item-related entities in a structured format. We aim to capture the spatio-temporal relationships of items and item-related entities by combining the modeling of sequential dependency (temporal relationships) and knowledge graphs (spatio relationships). We argue that leveraging knowledge graphs to make use of such additional information properly helps in making a better recommendation. Specifically, we extend 
 TSCR by leveraging knowledge graphs in the following ways: (a) we utilize knowledge graphs to train representations of items and item-related entities offline 
 and use them as initialized representations in the model;
and (b) given the rich structural information of knowledge graphs, we augment the user sequence of conversations with the multi-hop paths between two nodes (\ie items or item-related entities) in knowledge graphs and then train the model based on the augmented sequences. 

To sum up, our main contributions are as follows:

\begin{enumerate}[(a)]
    \item We propose a method, TSCR, which models the sequential dependencies in the conversations to improve \ac{CRS}.
    To the best of our knowledge, this is the first effort to explicitly model the bidirectional sequential dependency in natural language conversations for \acp{CRS}.
    \item We propose an extension model, TSCRKG, with a novel method to incorporate knowledge graphs in offline representation learning and knowledge graph-enhanced sequence modeling. 
    \item The extensive experiments on the two \ac{CRS} datasets demonstrate the effectiveness of our TSCR model, and our extension model TSCRKG significantly improves the recommendation performance compared to various state-of-the-art baselines and TSCR. 
\end{enumerate}
Among these contributions, (a) and part of (c) (namely the experiments for TSCR on the ReDial dataset) were covered in our prior work~\citep{zou2022improving-sigir}, (b) and part of (c) (namely the experiments for TSCRKG, the experiments for both models on the TG-ReDial dataset, and the case studies) are new contributions in this extended version.
 Specifically, the new contributions are in the following aspects.
 (1) First, we propose a new extension model, TSCRKG (See Sections~\ref{subsec:offinit} and~\ref{subsec:kgseq}). 
(2) Second, we conduct the experiments on an additional dataset, TG-ReDial, which demonstrates the effectiveness of TSCR and TSCRKG in the Chinese conversational recommendation scenario, besides the English scenario. The experiments on the additional TG-ReDial dataset yield some different insights and conclusions from the original ReDial dataset (See Section \ref{sec:exp}.). (3) Third, we form a research question and conduct an analysis to explore the impact of knowledge graph-enhanced representation learning and knowledge graph-enhanced sequence modeling (See Section \ref{subsec:impkg}.). (4) Forth, we add a case study to illustrate the ability of TSCR and TSCRKG to generate recommendations (See Section \ref{sec:case}.). 

The rest of this paper is organized as follows. In Section \ref{sec:rel}, we discuss the related work. In Section \ref{sec:meth}, we introduce our models TSCR and TSCRKG in detail. Section \ref{sec:exp} describes the conducted experiments and the corresponding analysis of experimental results. Section \ref{sec:conc} concludes the paper and summarizes some future work.

\section{Related Work}
\label{sec:rel}
In this section, we review the related work from the following two categories: conversational recommender system and knowledge graph learning. A large number of studies have been conducted on these topics. In this paper, we will review only the work that is most closely related to our research.

\subsection{Conversational Recommender System}
Thanks to the great power of collecting users' explicit feedback, conversations have shown great potential in a wide range of areas, such as information-seeking systems \citep{radlinski2017theoretical}, conversational product search \citep{zou2022learning}, dialog systems \citep{stoyanchev2014towards, de2003analysis}, and CRS \citep{SepliarskaiaKRR18, zou2022improving-sigir}. In information-seeking systems, early exploration investigated mixed-initiative systems by interacting with users via script-based conversation during a search session \citep{Belkin1995CasesSA}. Recently, researchers investigated conversational information-seeking systems by asking clarifying questions \citep{aliannejadi2019asking, zou2023asking}, to enhance the ability to understand the users' underlying information needs and retrieve the right information \citep{aliannejadi2019asking, HashemiZC20, radlinski2017theoretical, white2001questions, rosset2020leading}. In conversational product search, existing studies usually involve learning to ask strategies \citep{zou2022learning} to ask informative clarifying questions to form conversations \citep{zhang2018towards}, in order to locate relevant products a user is willing to purchase \citep{bi2019conversational}. Dialog systems usually structure conversations in multiple turns, including task-oriented systems \citep{madotto2020learning}, open-domain dialogs \citep{banchs2012iris}, and question-answering dialog systems \citep{de2003analysis}. \ac{CRS} is an emerging field that utilizes human-like natural language to deliver personalized and engaging recommendations through conversational interfaces like chatbots and intelligence assistants \citep{gao2021advances,jannach2021survey}. It is gaining considerable attention in recent years, driven by the rapid development of dialog systems. When considering CRS primarily as dialog systems, CRS can be regarded as either (1) system-driven (e.g., critiquing-based systems \citep{viappiani2007conversational}), (2) user-driven~\citep{loh2010recommendation}, or (3) mixed-initiative systems (most of CRS approaches) \citep{zou2022improving-sigir, loh2010recommendation}, based on who takes the initiative in the dialog \citep{jannach2021survey}. The dialog systems could generate flexible and contextual responses, however, they may suffer from meaningless responses and fail to drive the conversation toward more attractive states for making recommendations or suggestions. One type of solution tends to enable conversational systems the ability for proactive conversations \citep{liu2020towards, zhu2021proactive, deng2024towards} to alleviate this problem. Various aspects have been explored in this field, including proposing new intents and conversational slots through ontology expansion and actively analyzing failures in novel situations (e.g., \citep{lin2020discovering, liu2019proactive, zhang2021discovering}), learning to ask questions to move the conversation forward (e.g., \citep{ren2021learning, zou2022learning, zou2020towardsb}), leveraging task information and domain knowledge to guide purposeful topic shifts and exploration (e.g., \citep{liao2020topic, shah2021applying}), controlling the quality of response generation (e.g., \citep{baheti2021just, tang2021high}), and improving evaluation (e.g., \citep{tang2022re}) etc.

Early efforts in the development of CRS can be traced back to the work of  ~\citet{bridge2002towards, carenini2003towards, mahmood2009improving, felfernig2011developing}, and \citet{ thompson2004personalized}. More recently, various feedback mechanisms have been explored~\cite{chen2020balancing, zhao2013interactive, loepp2014choice, graus2015improving, christakopoulou2016towards, yu2019visual, sardella2019approach, jin2019musicbot, wu2019deep}.~\citet{zhao2013interactive} investigated interactive collaborative filtering, proposing methods to enhance probabilistic matrix factorization~\cite{mnih2008probabilistic} by using linear bandits to select items for user feedback.~\citet{loepp2014choice} explored set-based feedback, while ~\citet{graus2015improving} focused on choice-based feedback to learn latent factors and perform online interactive preference elicitation. We refer the reader to ~\citet{he2016interactive} and ~\citet{jugovac2017interacting} for a comprehensive review of interactive recommendations.

Recently, the research on \ac{CRS} has been classified into two categories generally~\citep{zhang2021kecrs,zhou2020towards}, including anchor based \acp{CRS} \citep{zou2020towardsb} and dialog based \acp{CRS} \citep{zou2022improving-sigir}.

Anchor-based based \acp{CRS}~\citep{zou2020towardsb} are in the form of ``system asks – user response'' mode. That is, systems could ask questions, then users provide feedback, and thus systems would use this information to refine their recommendations. For building effective anchor-based \acp{CRS}, recent studies have applied various ways for generating anchors to characterize items, including intent slots (e.g., item aspects and facets)\citep{zhang2018towards,sun2018conversational}, entities \citep{zou2020towardsb}, topics \citep{christakopoulou2018q}, and attributes \citep{lei2020estimation,lei2020interactive,hu2022learning,zhang2022multiple}. These anchors are usually collected to build a predefined question pool. Based on the constructed question pool, a line of work adopt multi-armed bandit \citep{christakopoulou2016towards, christakopoulou2018q, li2021seamlessly}, reinforcement learning \citep{lei2020estimation}, or greedy strategies \citep{christakopoulou2016towards, zou2022learning} to select appropriate questions to simulate multi-turn interactions to interact with users. By doing so, user feedback through conversational interactions is leveraged to optimize recommendation policies. Other than selecting appropriate questions, deep reinforcement learning is also applied to train a policy network to determine whether to recommend items or ask a question in a turn \citep{lei2020interactive, hu2022learning}. This category of studies usually relies on predefined dialog templates to construct conversations and focuses on how to offer recommendations within the shortest number of conversation turns \citep{zou2020towardsb}. In other words, they do not focus on modeling natural language conversations \citep{chen2019towards}. 

The other category is called dialog based \ac{CRS}~\citep{liu2020towards, zhou2020towards, liao2019deep, chen2021knowledge, liang2021learning, deng2022unified, pan2022keyword}, which is based on human-generated dialogs. Unlike anchor-based \acp{CRS} simulating conversations on the basis of extracted anchor text, dialog based \acp{CRS} focus on generating human-like responses while making accurate recommendations. At the early stage, as there are no publicly available datasets for \acp{CRS}, \citet{li2018towards} propose a \ac{CRS} dataset in the movie domain, named ReDial. Subsequently, \citet{zhou2020towards} propose the dataset TG-ReDial for \acp{CRS} in the Chinese language scenario. For datasets proposed in \ac{CRS} like ReDial and TG-ReDial, a human-like conversation usually contains items and entities, which can be linked to a knowledge base (e.g., DBpedia). Therefore, many existing studies incorporate external knowledge graphs for enriching the conversation information and capturing user preference accurately \citep{wang2021conversation, chen2019towards, zhou2020improving, zhang2021kecrs, zhang2021kers}. These studies typically employ a graph neural network to encode the knowledge graph and user preferences, along with a dialog management module to guide the conversation. For example, \citet{chen2019towards} introduce the entity-oriented knowledge graph, i.e., DBpedia \citep{bizer2009dbpedia}, to understand the user's intention. They adopt Relational Graph Convolutional Networks (R-GCNs) to learn entity representations in knowledge graphs for item recommendations and apply the Transformer framework to generate natural language responses. Based on the work of \citet{chen2019towards}, which mainly focuses on the entity-oriented knowledge graph, \citet{zhou2020improving} expand upon this by introducing an additional knowledge graph, i.e., the word-oriented knowledge graph -- ConceptNet \citep{speer2017conceptnet}. They propose a novel knowledge graph-based semantic fusion model for \ac{CRS} by utilizing the two knowledge graphs. Based on this, \citet{zhou2022c2} propose a contrastive learning approach from the coarse-to-fine perspective to improve the fusion of data semantics. In order to reduce the amount of redundant information in knowledge graphs, \citet{sarkar2020suggest} introduce different subgraphs to improve the performance of the recommendation for \acp{CRS}. To make use of multiple reasoning paths in knowledge graphs, \citet{CRWalker} and \citet{zhou2021crfr} propose \ac{CRS} models to perform multi-hop reasoning in knowledge graphs to track users' interest shifts. To alleviate the limitation of incomplete knowledge graphs, \citet{zhang2023variational} present a variational reasoning approach \citep{ren2022variational} with dynamic knowledge reasoning over incomplete knowledge graphs. In addition to knowledge graphs, pre-trained language models are also applied to enhance \acp{CRS} \citep{wang2022towards, wang2022recindial}. \citet{wang2022recindial} finetune the largescale pre-trained language models and propose a unified pre-trained language model-based framework to address the low-resource challenge in \acp{CRS}. Similarly, \citet{wang2022towards} apply a pre-trained language model and propose a \ac{CRS} model to unify the recommendation and conversation subtasks into the prompt learning paradigm. Another promising direction in conversational recommendation is the use of side information \citep{lu2021revcore, yang2021improving}. For instance, \citet{lu2021revcore} incorporate the review information of items to improve the recommendation performance in \acp{CRS}. \citet{li2022user} leverage multi-aspect user information to learn multi-aspect user preferences, while \citet{yang2021improving} introduce item meta information to improve item representations. 

This paper follows the line of the second category and focuses on the recommendation module only. Similar to the existing work using knowledge graphs, we also leverage the knowledge graph into our model but use it in both offline representation learning (i.e., we use the presentations of items and non-item entities learned based on the knowledge graph as the initialized embeddings in the recommendation module) and knowledge graph-enhanced sequence modeling (i.e., we augment the sequences of items and non-item entities in the conversations with paths from the knowledge graph and train the recommendation module based on both the original sequences and the augmented sequences). Moreover, different from the aforementioned studies, we explicitly model the bidirectional sequential dependency in natural language conversations for \acp{CRS}.

\begin{figure}[t]
\centering
\includegraphics[width=1\columnwidth]{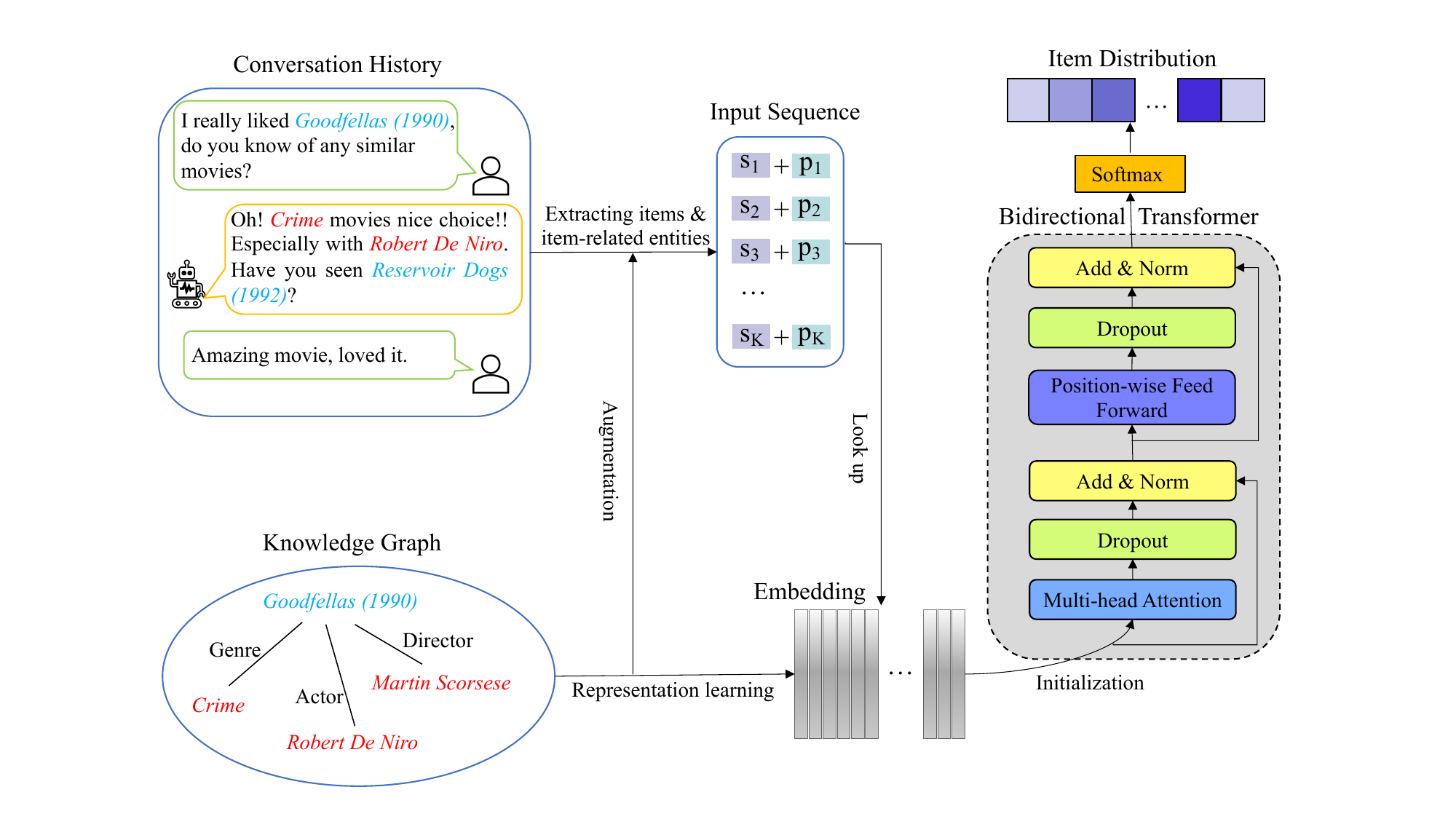}
\caption{The overview of our model. Our model extracts the items and item-related entities to form an input sequence, and then uses the knowledge graph and bidirectional Transformer to generate item recommendations.}
\label{fig:method}
\end{figure}
\subsection{Knowledge Graph Learning}
There has been a growing interest of research on knowledge graph learning. Knowledge graph-based representation learning learns the representations of entities in a knowledge graph in the form of high-dimensional vectors, while preserving the intrinsic properties of the knowledge graph. This learned representation can then be conveniently applied to various downstream tasks, such as question answering \citep{bordes2014question}, and recommender systems \citep{guo2020survey}. \citet{ji2021survey} present a survey on knowledge graph learning. 
For knowledge representation learning, various models have been proposed, such as TransE \citep{bordes2013translating}, TransH \citep{wang2014knowledge}, TransR \citep{lin2015learning} TransD \citep{lin2015learning}, HolE \citep{nickel2016holographic}, and R-GCN \citep{schlichtkrull2018modeling}. In this paper, we use R-GCN \citep{schlichtkrull2018modeling} to learn representations  of the entities (items and other non-item entities) in knowledge graphs. 

Moreover, several previous studies \citep{rezayi2021edge, kartsaklis2018mapping, wang2016text} have explored the integration of external text into knowledge graph embeddings. For example, 
\citet{sun2018open} employ Graph Convolutional Networks (GCN) \citep{kipf2016semi} to extract subgraphs that combine the semantics of both graphs and text sentences for question-answering tasks. \citet{malaviya2020commonsense} propose a transfer learning approach to enhance graph representations from pre-trained language models. \citet{rezayi2021edge} create augmented graphs and learn representations based on the original and augmented graphs. 

In this paper, we apply knowledge graphs to the specific downstream task of conversational recommendation. 
We utilize knowledge graphs for learning offline embeddings of items and non-item entities to initialize our model, and take advantage of the graph structure to enhance the user sequences in conversations. 

\begin{table}[tb]
\caption{{Notations.}}
\label{table:Notations}
\begin{center}
\begin{tabular}{l|p{0.75\columnwidth}}
\toprule
Notation & Explanation\\
\hline
$u$, $\mathcal{U}$ & a user, and a set of all users\\
$v$, $\mathcal{V}$ & an item, and a set of all items\\
$e$, $\mathcal{E}$ & an entity, and a set of all entities\\
$s_k$, $\mathcal{S}$ & the $k$-th element (entity) in a user sequence, and a user sequence of a conversation\\
$\mathcal{S}^{\prime}_u$, $\mathcal{S}^{(mask)}_u$, $\mathcal{S}_u^{*}$ & the masked version for user historical sequence $\mathcal{S}_u$, the set of masked items in $\mathcal{S}_u$, and the augmented sequence by the knowledge graph\\
$\mathbf{s}_k$ & the entity embedding of the $k$-th  element of a user sequence\\
$\mathbf{p}_k$ & the position embedding of the $k$-th element of a user sequence\\
$\mathbf{h}^n$, $\mathbf{H}^n$ & the hidden embedding of an element in a user sequence in the $n$-th layer of Transformer, and the trainable embedding matrix of all elements in the $n$-th layer of Transformer\\
$\mathcal{G}$ & a knowledge graph \\
$r$, $\mathcal{R}$ & a relation, and a relation set\\
$\mathbf{n}_{e}^{(l)}$  & the representation of node (entity) $e$ at the $l$-th layer of R-GCN\\
$\mathbf{W}$ & a learnable transformation matrix in R-GCN\\
\bottomrule
\end{tabular}
\end{center}
\end{table}
\section{Sequential Conversational Recommender System}
\label{sec:meth}
We suppose there is a user set $\mathcal{U}$ = $\{u_1, u_2, \dots, u_{|\mathcal{U}|}\}$, item set $\mathcal{V}$ = $\{v_1, v_2, \dots, v_{|\mathcal{V}|}\}$, and conversations $\mathcal{D}$. We extract entities $\mathcal{E}$ = $\{e_1, e_2, \dots, e_{|\mathcal{E}|}\}$ from conversations $\mathcal{D}$ based on DBpedia. The entity set $\mathcal{E}$ consists of all the items and other item-related entities (i.e., $\mathcal{V} \subseteq \mathcal{E}$). For a user $u\in \mathcal{U}$, we have his/her entity mention (i.e., items or other item-related entities) history, extracted from his/her conversation, denoted as a sequence $\mathcal{S}_u$ = $[s^u_1, \dots, s^u_k, \dots, s^u_K]$ ($s^u_k \in \mathcal{E}$), we aim to accurately predict the next item $v^*$ that user $u$ likes, along the development of the conversation.   
The main notations used throughout the paper are summarized in Table~\ref{table:Notations}.

In the following, we first describe the TSCR model \citep{zou2022improving-sigir}, in particular the base model adopted, \ie Transformer~\citep{vaswani2017attention, sun2019bert4rec} (Section~\ref{subsec:transformer}), and how we train our model and perform the item recommendation (Section~\ref{subsec:prediction}). Then, we express in detail how we incorporate the knowledge graph for our extension model TSCRKG, in particular, (a) knowledge graph enhanced representation learning (Section~\ref{subsec:offinit}) and (b) knowledge graph enhanced sequence modeling (Section~\ref{subsec:kgseq}). 

The overview of our model is shown in Figure~\ref{fig:method}. The TSCR model extracts the items and item-related entities to form an input sequence, and then uses the bidirectional Transformer as the base model (Section~\ref{subsec:transformer}) to generate item recommendations by applying a Cloze task (Section~\ref{subsec:prediction}). The TSCRKG model extends the TSCR model by leveraging the knowledge graph for offline representation learning (Section~\ref{subsec:offinit}) and sequence augmentation (Section~\ref{subsec:kgseq}). 

\subsection{Base Model}
\label{subsec:transformer}
Inspired by \citet{sun2019bert4rec}, 
we adopt Transformer~\citep{vaswani2017attention, sun2019bert4rec} %
as our base model, which consists of the embedding layer, self-attention layer, and prediction layer.

\paragraph{Embedding layer.} Given a sequence, we denote the embedding for the element at position $k$ in the input sequence as $\mathbf{h}^0_k$. 
For the representation of $\mathbf{h}^0_k$, we inject a learnable position embedding, $\mathbf{p}_k$, into the embedding of each element of the input sequence, $\mathbf{s}_k$: 
\begin{equation}
\label{equ:4}
\mathbf{h}^0_k= \mathbf{s}_k + \mathbf{p}_k.
\end{equation}

All elements together form a trainable embedding matrix $\mathbf{H}^0$. Based on this initially trainable embedding matrix 
$\mathbf{H}^0$, we interactively calculate $\mathbf{H}^n$ at each Transformer layer $n$. 

\paragraph{Self-attention layer.} A self-attention layer consists of two sub-layers: a multi-head self-attention sub-layer and a Position-wise Feed-Forward Network (PFFN). More details can be found in \citet{vaswani2017attention}.
\begin{equation}
\label{equ:1}
\mathbf{H}^{n+1}=\text{MultiHead}(\text{PFFN}(\mathbf{H}^{n})).
\end{equation}

We construct the PFFN by the Feed-Forward Network (FFN) with GELU activation \citep{hendrycks2016bridging} at each position separately:
\begin{equation}
\label{equ:2}
\text{PFFN}(\mathbf{H}^{n})=[\text{FFN}(\mathbf{h}^{n}_1)^\intercal;\dots;\text{FFN}(\mathbf{h}^{n}_k)^\intercal]^\intercal.
\end{equation}

In addition, we deploy a residual connection around each of the two sub-layers, followed by a dropout and layer normalization, \ie the output of each sub-layer is
actually: 
\begin{equation}
\label{equ:3}
\text{LayerNorm}(\mathbf{H}^{n} + \text{Dropout}(\text{sublayer}(\mathbf{H}^{n})),
\end{equation}
where sub-layer is MultiHead or PFFN in Eq. \ref{equ:1}.

\paragraph{Prediction layer}
After N layers of Transformer, we get the final output $\mathbf{H}^{N}$ for the input sequence. Assuming we mask $s_k$ at the input sequence, we then utilize $\mathbf{h}^{N}_k$ to predict the masked item $s_k$. Specifically, we apply a softmax function through a two-layer FFN with GELU activation in between to produce an output distribution over items. To ensure recommendations are all items, we set the score of non-item entities in the softmax function to $-\infty$. 

In this work, we adopt the Transformer architecture, which utilizes attention mechanisms to capture temporal relations while processing input tokens of a sequence in parallel. The Transformer architecture has been demonstrated as a powerful framework for supporting large-scale training datasets with enough parameters \cite{lin2022survey}. Notably, the Transformer architecture makes few prior assumptions about the structural information of data but does not make any assumptions about how the data is structured. This makes Transformer a universal and flexible architecture that is well-suited for capturing dependencies across various ranges. However, this also poses a challenge: making Transformer difficult to train on small-scale datasets. The methods to alleviate this issue are to conduct pre-training on extensive unlabeled datasets and introduce structural bias or regularization into the model \cite{guo2019low}. Another key challenge of applying Transformer lies in its inefficiency when processing long sequences, primarily due to the computation and memory complexity of the self-attention module. The self-attention's complexity concerning sequence length can significantly constrain the performance of downstream tasks. To overcome this challenge, one can apply improvement methods including the incorporation of lightweight attention (e.g., sparse attention variants \cite{child2019generating}) and the application of divide-and-conquer strategies (decomposing an input sequence into finer segments, facilitating more efficient processing by Transformer or Transformer modules), such as recurrent and hierarchical Transformers \cite{lin2022survey}. 

\subsection{Masked Item Prediction}
\label{subsec:prediction}
We apply a Cloze task \citep{devlin2018bert, taylor1953cloze} on the sequence from the conversational history of a user  (\ie the sequence of mentioned items and item-related entities) to train our model. Given a sequence $\mathcal{S}_u$, we randomly mask a proportion of items 
in the input sequence by replacing them with the special token ``[mask]'', and then predict the original IDs of the masked items (in our implementation, items are represented by their corresponding item IDs). 
Following BERT \citep{devlin2018bert}, we leverage the bidirectional contextual information in the input sequence for predicting the masked item. We use the negative log-likelihood of the masked targets as the loss:
\begin{equation}
\label{equ:6}
\mathcal{L} = \frac{1}{|\mathcal{S}^{(mask)}_u|} \sum_{v^{\prime} \in \mathcal{S}^{(mask)}_u} - \log P(v^{\prime}|\mathcal{S}^{\prime}_u),
\end{equation}
where $\mathcal{S}^{\prime}_u$ is the masked version for user historical sequence $\mathcal{S}_u$, $\mathcal{S}^{(mask)}_u$ is the set of masked items in $\mathcal{S}_u$, and $v^{\prime}$ is one of the masked items.

For testing, it is not practical to use bidirectional information to predict as the testing item is always in the future given the current context. To this end, we construct a contextual sequence for each testing item, and then add a ``[mask]'' token to the end of the sequence to predict a testing item. For example, if there are three items in a sequence, we mask the first item and predict it with possible entities that are already mentioned in the dialog till this prediction. Then we mask and predict the second item based on the first item and other item-related entities mentioned so far in the dialog till this prediction, then the third item. To better match the last item prediction during testing, we also mask the last item for each training sequence to generate a training sample during training.  

Contrary to bidirectional Transformer models like BERT~\citep{vaswani2017attention}, which is a pre-training model for sentence representation, our model is an end-to-end model trained for sequential conversational recommendation. Also, we removed the next sentence loss and next sentence prediction since there is only one sequence of user's historical mentions of items and other item-related entities in \ac{CRS}. Different from those studies using bidirectional Transformer or pre-trained BERT for recommender systems, we trained our model in an end-to-end style and incorporated the conversational information (\eg entities), aiming to improve \ac{CRS}. 

\begin{algorithm}[tb]
\SetKwInOut{Input}{input}
\SetKwInOut{Output}{output}
\Input{A user historical sequence $\mathcal{S}_u$, the length of the user sequence $K$, a knowledge graph $\mathcal{G}$}
\Output{An augmented sequence $\mathcal{S}_u^{*}$}
$\mathcal{S}_u = [s^u_1, \dots, s^u_k, \dots, s^u_K]$: $K>1$\\
Pair each of two neighboring entities in $\mathcal{S}_u$:\\
\qquad $\mathcal{D}_s = \{(s_1,s_2), \dots, (s_k,s_{k+1}), \dots, (s_{K-1},s_{K})\}$\\
Set $\mathcal{S}_u^{*}$ as an empty sequence\\
$k \leftarrow 1$\\
\While{$k < K$}{
For the pair $(s_k,s_{k+1}) \in \mathcal{D}_s$, find the shortest path from $s_k$ to $s_{k+1}$ in $\mathcal{G}$ by A-Star:\\
\qquad $\mathcal{N} = \text{A-Star}(s_k,s_{k+1})$\\
Augment $\mathcal{S}_u^{*}$ by adding entities in the shortest path sequentially: \\
\qquad $\mathcal{S}_u^{*} = [\mathcal{S}_u^{*}; \mathcal{N}]$\\
$k \leftarrow k + 1$\\
} 
\caption{Sequence augmentation by the knowledge graph}
\label{algo1}
\end{algorithm}

\subsection{Offline Representation Learning}
\label{subsec:offinit}
We introduce an offline representation learning technique that utilizes an external knowledge graph to initialize our model TSCRKG. Given it is difficult to comprehensively understand user preferences based solely on conversational context, the inclusion of external knowledge is necessary to encode user preferences. For dialogs in \acp{CRS}, item mentions and item-related entities can be extracted to construct external knowledge graphs. Inspired by the previous studies \citep{chen2019towards, zhou2020improving}, we introduce a
knowledge graph sourced from DBpedia \citep{lehmann2015dbpedia} to encode structural and relational information in the knowledge graph. Specifically, we perform entity linking \citep{ferragina2010tagme} to map the item mentions and item-related entities in the dataset to DBpedia. With the help of the external knowledge graph, it enables us to model user preferences more accurately. 

Given a knowledge graph $\mathcal{G}$ (i.e., DBpedia), it consists of an entity set $\mathcal{E}$ and a relation set $\mathcal{R}$. The knowledge graph $\mathcal{G}$ stores semantic facts in the form of a triple $<e_1, r, e_2>$, where $e_1, e_2 \in \mathcal{E}$ are items or item-related entities from the entity set $\mathcal{E}$ and $r \in \mathcal{R}$ is the relation between $e_1$ and $e_2$ from the relation set $\mathcal{R}$. 

In this paper, we employ R-GCN \citep{schlichtkrull2018modeling} to encode entity representations in the knowledge graph $\mathcal{G}$. Specifically, we pre-train the representation $e \in \mathcal{E}$ by using R-GCN to initialize the offline embeddings of items or other item-related entities. Formally, in R-GCN, the representation of node $e$ at $(l + 1)$-th layer is calculated as:
\begin{equation}
\label{equ:rgcn}
\mathbf{n}_{e}^{(l+1)}=\sigma(\sum_{r \in \mathcal{R}} \sum_{e^{\prime} \in \mathcal{E}_{e}^{r}} \frac{1}{Z_{e, r}} \mathbf{W}_{r}^{(l)} \mathbf{n}_{e^{\prime}}^{(l)}+\mathbf{W}^{(l)} \mathbf{n}_{e}^{(l)}),
\end{equation}
where $\mathbf{n}_{e}^{(l)}$ denotes the representation of node (i.e., entity) $e$ at the $l$-th layer, and $\mathcal{E}_{e}^{r}$ is the set of neighboring nodes for $e$ under the relation $r$. $\mathbf{W}^{(l)}$ is a learnable transformation matrix for transforming the representations of nodes at the $l$-th layer $\mathbf{n}_{e}^{(l)}$, while $\mathbf{W}_{r}^{(l)}$ is another learnable relation-specific matrix for transforming the embedding of neighboring nodes with the relation $r$. $Z_{e, r}$ is a normalization factor. 

After aggregating the structural and relational information of the knowledge graph, we obtain all the node representations on the top R-GCN layer. These knowledge-enhanced node representations are then used to initialize $\mathbf{s}_k$ in Equation \ref{equ:4}, i.e., the embeddings of items or other item-related entities in the dataset, for the Transformer model. 

\subsection{Knowledge Graph Enhanced Sequence Modelling}
\label{subsec:kgseq}
In TSCRKG, we utilize the knowledge graph to enhance the user sequence of conversations (i.e., the sequence of mentioned items and item-related entities in the conversation). Specifically, we introduce an approach to augment the user sequence with the multi-hop paths in the knowledge graph. 

Given a user historical sequence $\mathcal{S}_u$, we aim to generate an augmented sequence $\mathcal{S}_u^{*}$, which is able to maintain the useful structural information of the knowledge graph but also inject the global collaborative context into the augmented sequence to facilitate the sequence modeling. The algorithm for sequence augmentation by the knowledge graph is shown in Algorithm~\ref{algo1}. Let $\mathcal{S}_u$ = $[s^u_1, \dots, s^u_k, \dots, s^u_K]$, we first pair each of two neighboring entities in the user sequence $\mathcal{S}_u$. For each pair, we find the shortest path in the knowledge graph (if any) between the two entities, and then augment the user sequence by sequentially adding all the entities in the shortest path following the last entity of the user sequence. Take the pair $(s^u_1, s^u_2)$ in the sequence as an example, if there are paths existing in the knowledge graph between $s^u_1$ and $s^u_2$, we augment the user sequence by filling the entities in the shortest path between $s^u_1$ and $s^u_2$. If there is a shortest path $s^u_1 \rightarrow s_k \rightarrow s^u_2$, then $[\dots, s^u_1, s^u_2, \dots]$ in the original user sequence $\mathcal{S}_u$ will replaced by $[\dots, s^u_1, s_k, s^u_2, \dots]$ in $\mathcal{S}_u^{*}$. If there is no path existing in the knowledge graph between $s^u_1$ and $s^u_2$, then $[\dots, s^u_1, s^u_2, \dots]$ in the user sequence will still keep the same. In this way, (a) we complete the sequence in which some entities in the path are missing. This enables us to alleviate the problem of incomplete user sequences; (b) this helps us to have a complete path for knowledge graph reasoning, which improves the explainability of the recommendation; (c) this data augmentation allows us to increase the number of training data. For extracting the shortest path between two entities in the knowledge graph, we apply $\text{A}^*$(A-Star) algorithm \citep{saian2016optimized}, which is one of the best-known graph searching algorithms to find the shortest paths from the initial node to final node \citep{saian2016optimized}. 
It utilizes heuristic functions to guide its search, which is conducted by visiting nodes in the tree. The algorithm uses a best-first search and creates a path with the least cost through an evaluation function:
the algorithm selects the node with the lowest total cost by an evaluation function: 
\begin{equation}
\label{equ:AStar}
f(x) = g(x) + h(x),
\end{equation}
where $g(x)$ represents the distance so far to reach node $x$ (entity in this paper), $h(x)$ is a heuristic function to calculate the estimated distance from $x$ to target. $f(x)$ is the addition of $g(x)$ and $h(x)$, representing the estimated total cost of the path through $x$ to the target. The lowest value of $f(x)$ is determined for the shortest path. After the enhanced user sequence is obtained, we feed the enhanced sequence into the bidirectional Transformer (see Section \ref{subsec:transformer}) and apply a Cloze task to perform the masked item prediction (see Section \ref{subsec:prediction}), same with the TSCR model.

\section{Experiments and Analysis}
\label{sec:exp}
In this section, we evaluate our proposed models: TSCR and  TSCRKG. We first introduce our experimental settings, including the used datasets, evaluation metrics, parameter settings, and baselines. Then we report and analyze our experimental results. Through experiments, we aim to answer the following research questions:

\RQ{1}{How effective is our proposed simple model compared to current state-of-the-art baselines?}
\RQ{2}{What are the impacts of different knowledge graph components?}
\RQ{3}{What are the contributions of items and entities in a sequence?}
\RQ{4}{What is the effect of item position?}
\RQ{5}{How do the parameters of our proposed model affect its efficacy?}

RQ1, and RQ3 -- RQ5 were also investigated by \citet{zou2022improving-sigir}. In this paper, we extend \citet{zou2022improving-sigir} on answering RQ1 and RQ3 -- RQ5 by adding (1) more state-of-the-art baselines, (2) an additional dataset TG-ReDial for validating the effectiveness of TSCR in the Chinese \ac{CRS} scenario, and (3) additional experiments for the extension model TSCRKG. Also, we extend \citet{zou2022improving-sigir} by adding RQ2 to explore the impacts of knowledge graph representation learning and knowledge graph enhanced sequences. In addition, we add a case study in this paper to illustrate the ability of TSCR and TSCRKG to generate recommendations.

\begin{table}[t]
\caption{Statistics of the used datasets in our experiments.}
\label{table:data}
\centering
\begin{tabular}{lcccc}
\toprule
\textbf{Dataset} & \textbf{\# Conversations} & \textbf{\# Utterances} & \textbf{\# Users} & \textbf{\# Items}\\
\midrule 
ReDial & 10,006 & 182,150 & 956 & 51,699\\
TG-ReDial & 10,000 & 129,392 & 1,482 & 33,834\\
\bottomrule
\end{tabular}
\end{table}
\subsection{Experimental Setting}
\label{subsec:exp}
\subsubsection{Dataset} 
In this work, we conduct the experiments on two existing conversational recommendation benchmark datasets, ReDial and TG-ReDial, as done in \citet{ren2022variational} and  \citet{zhou2022c2}. The statistics of our datasets are shown in Table \ref{table:data}. 
\begin{itemize}
    \item \textbf{ReDial dataset}. Same as \citet{li2018towards, chen2019towards, sarkar2020suggest, zhou2020improving}, and~\citet{CRWalker}, we use the \textit{REcommendations through DIALog (ReDial)} to evaluate our model. ReDial is a English conversational recommendation dataset, including a set of annotated dialogs in which a seeker requests movie suggestions from the recommender. It contains 956 users, 51,699 movies, 10,006 conversations, and 182,150 utterances. 
    \item \textbf{TG-ReDial dataset}. The TG-ReDial dataset \citep{zhou2020towards} is a Chinese conversational recommendation dataset which is also 
    used in \citet{zhou2020towards, ren2022variational} and \citet{zhou2022c2}. The dialogs are generated between a user and a recommender in the movie domain via a topic-guided way. It contains 1,482 users, 33,834 movies, 10,000 dialogs, and 129,392 utterances. 
\end{itemize}
The datasets are split into training, validation, and test sets by 8:1:1 ratio. Besides movies (\ie the items), we extract the relevant entities, such as director and genre, from DBpedia, as suggested by~\citet{chen2019towards, sarkar2020suggest, zhou2020improving}, and~\citet{CRWalker}.

\subsubsection{Evaluation Metrics}
Following \citet{chen2019towards, zhou2020improving, zhou2022c2, lu2021revcore}, and \citet{zhang2023variational}, we use Recall@$k$ ($k$ = 1, 10, and 50) as our evaluation metrics for the recommendation task in \acp{CRS}. Recall@$k$ evaluates whether the target item provided by human recommenders appears in the top-$k$ items produced by the recommender system. 
Moreover, we use the Mean Reciprocal Rank (MRR) to indicate the mean of the reciprocal of the rank of the target item in the ranked list predicted by the model. 
For each conversation, we start from the first item (movie) to recommend in the recommender's responses. This means, each item in the recommender's responses is regarded as ground truth and we evaluate them one by one throughout the conversation following the previous work~\citep{chen2019towards, zhou2020improving}. For each testing instance, we rank all possible items within the dataset. 

\subsubsection{Parameter Settings} 
We train our model using Adam  \citep{kingma2014adam} and TensorFlow with a learning rate of 1e-4. We set the batch size = 256, the number of Transformer layers N = 2, head number = 2, the maximum sequence length K = 100, L2 regularization strength = 0.01, and the global norm clip of gradients = 5 for stable training. The number of R-GCN layers and the normalization factor $Z_{e, r}$ of R-GCN are set to 1. 
We study the effect of the hidden dimensionality and mask proportion in Section \ref{Para}. The hidden dimensionality ranges within [32, 64, 128, 256] and the mask proportion is tuned within the range of [0.2, 0.4, 0.6, 0.8]. For the parameter settings of all baselines, we use the results of each baseline under its optimal hyperparameter settings. 

\subsubsection{Baseline}
In this work, we consider two classical baselines and several strong baselines used against ReDial and TG-ReDial: 
\begin{itemize}
    \item \textbf{Popularity} is a classical baseline sorting the items according to historical recommendation frequency. 
    \item \textbf{TextCNN} is a classical CNN-based recommendation model learning embeddings from contextual utterances. 
    \item \textbf{ReDial} \citep{li2018towards} is the benchmark model of ReDial applying an autoencoder recommender for conversational recommendation. 
    \item \textbf{KBRD} \citep{chen2019towards} utilizes the knowledge graph of DBpedia to introduce knowledge-grounded information to improve conversational recommendation. 
    \item \textbf{KGSF} \citep{zhou2020improving} incorporates a knowledge graph enhanced recommender by utilizing both entity-oriented and word-oriented knowledge graphs for the conversational recommendation. 
    \item \textbf{KECRS}~\citep{zhang2021kecrs} is a knowledge-enriched conversational recommendation model based on a constructed knowledge graph in the movie domain called TMDKG.
    \item \textbf{RevCore}~\citep{lu2021revcore} is a review-augmented conversational recommender by incorporating reviews to enrich item information. 
    \item \textbf{$\text{C}^2$-CRS}~\citep{zhou2022c2} is one of the state-of-the-art conversational recommendation models based on coarse-to-fine contrastive learning framework and semantic fusion of multi-type data. For a fair comparison, we removed the review information from the $\text{C}^2$-CRS.
    \item \textbf{VRICR}~\citep{zhang2023variational} is one of the state-of-the-art conversational recommendation models, which is based on a variational reasoning approach over incomplete knowledge graphs. 
\end{itemize}
Among these baselines, Popularity, and TextCNN are classical recommendation methods, while ReDial, KBRD, KGSF, KECRS, RevCore, $\text{C}^2$-CRS, and VRICR are conversational recommendation methods. For parameters used in baselines, we utilize the optimal parameters reported in the corresponding paper.

\begin{table}[t]
\caption{Recommendation performances of our model TSCRKG, TSCR, and baselines on ReDial. `*' indicates significant improvements upon the best baseline in Fisher random test with $p$-value $<0.05$. Best performances are in bold. }
\label{table:1}
\centering
\begin{tabular}{lccc}
\toprule
\textbf{Model} & \textbf{Recall@1} & \textbf{Recall@10} & \textbf{Recall@50} \\
\midrule 
Popularity & 0.012 & 0.061 & 0.179\\
TextCNN & 0.013 & 0.068 & 0.191\\
ReDial & 0.023 & 0.129 &0.287\\
KBRD & 0.030 & 0.164 & 0.338\\
KGSF & 0.039 & 0.183 & 0.378\\
KECRS & 0.021 & 0.143 & 0.340\\
RevCore & 0.046 & 0.220 & 0.396\\
$\text{C}^2$-CRS & 0.050 & 0.218 & 0.395\\
VRICR & 0.057 & 0.251 & 0.416\\
\midrule
TSCR & 0.072* & 0.257* & 0.447* \\
TSCRKG & \textbf{0.087*} & \textbf{0.268*} & \textbf{0.465*} \\
\bottomrule
\end{tabular}
\end{table}

\begin{table}[t]
\caption{Recommendation performances of our model TSCRKG, TSCR, and baselines on TG-ReDial. `*' indicates significant improvements upon the best baseline in Fisher random test with $p$-value $<0.05$. Best performances are in bold. }
\label{tab:TGReDialPerfromance}
\centering
\begin{tabular}{lccc}
\toprule
\textbf{Model} & \textbf{Recall@1} & \textbf{Recall@10} & \textbf{Recall@50}\\
\midrule 
Popularity & 0.000 & 0.003 & 0.014 \\
TextCNN & 0.003 & 0.010 & 0.024 \\
ReDial & 0.000 & 0.002 & 0.013 \\
KBRD & 0.005 & 0.032 & 0.077 \\
KGSF & 0.005 & 0.030 & 0.074 \\
KECRS & 0.002 & 0.026 & 0.069 \\
RevCore & 0.004 & 0.029 & 0.075 \\
$\text{C}^2$-CRS & 0.005 & 0.031 & 0.077\\
VRICR & 0.005 & 0.032 & 0.081 \\
\midrule
TSCR & 0.005 & 0.032 & 0.080\\
TSCRKG & \textbf{0.007*} & \textbf{0.037*} & \textbf{0.090*} \\
\bottomrule
\end{tabular}
\end{table}
\subsection{Overall Performance (RQ1)}
In this section, we study how effective is our proposed method compared to prior solutions. We compare our recommendation performances with baselines on ReDial and TG-ReDial, as shown in Table~\ref{table:1} and Table~\ref{tab:TGReDialPerfromance}, respectively. The evaluation metrics are reported as the average performance for the max number of conversational turns. 

For the recommendation performance on the ReDial dataset from Table~\ref{table:1}, we observe that ReDial outperforms the classical recommendation models, Popularity and TextCNN, by using mentioned items in the dialog to make recommendations. However, for the recommendation performance on the TG-ReDial dataset from Table~\ref{tab:TGReDialPerfromance}, we observe that the ReDial model performs worse than Popularity and TextCNN. That is, ReDial performs much worse on the TG-ReDial dataset than the ReDial dataset. The possible reason is that the ReDial model relies heavily on the mentioned items in conversations, but the mentioned items in the TG-ReDial dataset are much sparser than those in the ReDial dataset. Furthermore, KBRD, KGSF, KECRS, RevCore, $\text{C}^2$-CRS, and VRICR outperform ReDial in both Table~\ref{table:1} and Table~\ref{tab:TGReDialPerfromance}, which might be because they introduce external knowledge graphs and entities to understand the user's intentions. 
Also, we see that our proposed model, TSCR, significantly outperforms all the baselines on all three metrics on the ReDial dataset. Take Recall@50 as an example, TSCR outperforms Popularity, TextCNN, ReDial, KBRD, KGSF, KECRS, RevCore, $\text{C}^2$-CRS, and VRICR by 150\%, 134\%, 56\%, 32\%, 18\%, 31\%, 13\%, 13\%, and 7\%, respectively.  
As for the TG-ReDial dataset, TSCR outperforms Popularity, TextCNN, ReDial, KBRD, KGSF, KECRS, RevCore, and $\text{C}^2$-CRS on all three metrics. TSCR achieves a similar performance with VRICR, which might be because that VRICR benefits from the leverage of the knowledge graph although it neglects the modeling of the sequential dependencies. The results indicate that our TSCR model is effective and incorporating the sequential occurrence of items and entities is highly beneficial for improving recommender performance in \ac{CRS}. But note that this does not mean KBRD, KGSF, KECRS, RevCore, $\text{C}^2$-CRS, and VRICR are worse than our TSCR model, as they focus more on the natural language response generation part and need to balance the recommender part and natural language response generation part by jointly modeling them in \ac{CRS}. 
In addition, different from most prior solutions which use knowledge graphs to reduce candidate item space~\citep{sarkar2020suggest}, TSCR does not make use of the structure of knowledge graphs, although a sequence in this work can be mapped to a path in knowledge graphs. In other words, TSCR only uses the knowledge base as a dictionary to extract the relevant entities for the items mentioned in a dialog. 

Moreover, we find that our proposed extension model, TSCRKG, significantly outperforms all the baselines on all metrics on both ReDial and TG-ReDial datasets. As for Recall@50 on the ReDial dataset, TSCRKG outperforms ReDial, KBRD, KGSF, KECRS, RevCore, $\text{C}^2$-CRS, and VRICR by 62\%, 38\%, 23\%, 37\%, 17\%, 18\%, and 12\% respectively. As for Recall@50 on the TG-ReDial dataset, TSCRKG outperforms ReDial, KBRD, KGSF, KECRS, RevCore, $\text{C}^2$-CRS, and VRICR by 592\%, 17\%, 22\%, 30\%, 20\%, 17\%, and 11\% respectively. While TSCR achieves similar performance with the state-of-the-art baseline VRICR on the TG-ReDial dataset, TSCRKG achieves a significant improvement over VRICR on the TG-ReDial dataset. This indicates that TSCRKG enhanced by the knowledge graph is more powerful than TSCR. Furthermore, TSCRKG achieves a significant improvement over TSCR on both ReDial and TG-ReDial datasets. TSCRKG outperforms TSCR by 21\% on Recall@1, 4\% on Recall@10, and 4\% on Recall@50, on the ReDial dataset, while outperforming TSCR by 40\% on Recall@1, 16\% on Recall@10, and 13\% on Recall@50, on the TG-ReDial dataset. This again demonstrates that TSCRKG powered by the structure of the knowledge graph is able to provide more accurate recommendations than TSCR. Learning offline representations from knowledge graphs and enhancing sequences by multi-hop paths of knowledge graphs are highly beneficial for item recommendations in \ac{CRS}.

\begin{table}[t]
\caption{The performance of TSCRKG on ReDial with different components removed. `*' indicates significant improvements upon TSCRKG -w/o both in Fisher random test with $p$-value $<0.05$. Best performances are in bold. }
\label{tab:TSCRKG_KGablation}
\centering
\begin{tabular}{lcccc}
\toprule
\textbf{Model} & \textbf{Recall@1} & \textbf{Recall@10} & \textbf{Recall@50} \\
\midrule 
TSCRKG & \textbf{0.087*} & \textbf{0.268*} & \textbf{0.465*} \\
\midrule
-w/o KGseq & 0.085* & 0.266* & 0.449* \\
-w/o offline & 0.080* & 0.259* & 0.447 \\
-w/o both & 0.072 & 0.257 & 0.447 \\%
\bottomrule
\end{tabular}
\end{table}

\begin{table}[t]
\caption{The performance of TSCRKG on TG-ReDial with different components removed. `*' indicates significant improvements upon TSCRKG -w/o both in Fisher random test with $p$-value $<0.05$. Best performances are in bold. }
\label{tab:TSCRKG_KGablation2}
\centering
\begin{tabular}{lccc}
\toprule
\textbf{Model} & \textbf{Recall@1} & \textbf{Recall@10} & \textbf{Recall@50} \\
\midrule 
TSCRKG & \textbf{0.007*} & \textbf{0.037*} & \textbf{0.090*} \\
\midrule
-w/o KGseq & 0.006* & 0.034* & 0.084* \\
-w/o offline & 0.007* & 0.034* & 0.086* \\
-w/o both & 0.005 & 0.032 & 0.080 \\
\bottomrule
\end{tabular}
\end{table}
\subsection{Impact of Knowledge Graph Representation Learning and Knowledge Graph Enhanced Sequences (RQ2)}
\label{subsec:impkg}
Next, we answer RQ2. To understand what are the contributions of different knowledge graph components, 
we conduct an ablation study comparing our model with its ablation variants (TSCRKG removing the offline representation learning on knowledge graphs, represented by ``-w/o offline'', TSCRKG removing the part of knowledge graph enhanced sequences, represented by ``-w/o KGseq'', and TSCRKG removing both offline representation learning and knowledge graph enhanced sequences, represented by ``-w/o both''). 
The results on the ReDial dataset are shown in Table~\ref{tab:TSCRKG_KGablation} and the results on the TG-ReDial dataset are shown in Table~\ref{tab:TSCRKG_KGablation2}. From what we observed in Table~\ref{tab:TSCRKG_KGablation} and Table~\ref{tab:TSCRKG_KGablation2}, either removing the offline representation learning or removing path-enhanced sequences over knowledge graphs lowers the recommendation performance on all evaluation metrics. 
Specifically, on the ReDial dataset, the performance of TSCRKG -w/o KGseq, TSCRKG -w/o offline, and TSCRKG -w/o both drops by 3\% and 4\% and 4\% on Recall@50, respectively. On the TG-ReDial dataset, the performance of TSCRKG -w/o KGseq, TSCRKG -w/o offline, and TSCRKG -w/o both drops by 7\% and 4\% and 11\% on Recall@50, respectively. 
This demonstrates that both the two knowledge components play an essential role in item recommendations. Moreover, offline representation learning is more helpful for item recommendations on the ReDial dataset because the performance drops more without offline representation learning than without knowledge graph enhanced sequences on the ReDial dataset. In contrast, the knowledge graph enhanced sequences are more helpful on the TG-ReDial dataset since TSCRKG -w/o KGseq achieves lower performance than TSCRKG -w/o offline on the TG-ReDial dataset. This indicates that offline representation learning is the more important component on the ReDial dataset while the knowledge graph enhanced sequence is the more important component on the TG-ReDial dataset. This discrepancy might be because that item mentions are much more sparse and the user sequences in conversations are shorter in the TG-ReDial dataset, leading to more need to enhance user sequences by the knowledge graph. This suggests that, when the data suffers from the lack of contextual information in conversations, future research on \ac{CRS} models can incorporate the structure of external knowledge graphs to increase the number of effective entities, instead of simply mapping mentioned entities to knowledge graphs.

\begin{table}[t]
\caption{Recommendation performance comparison of our model TSCR, TSCR w/o entity, and TSCR w/o item on ReDial. `*' indicates significant improvements upon the best baseline in Fisher random test with $p$-value $<0.05$. Best performances are in bold. }
\label{tab:TSCR_seqablation}
\centering
\begin{tabular}{lccc}
\toprule
\textbf{Model} & \textbf{Recall@1} & \textbf{Recall@10} & \textbf{Recall@50} \\
\midrule 
TSCR & \textbf{0.072*} & \textbf{0.257*} & \textbf{0.447*} \\
\midrule
-w/o entity & 0.070& 0.251& 0.428 \\
-w/o item & 0.033 & 0.148 & 0.320 \\
\bottomrule
\end{tabular}
\end{table}

\begin{table}[t]
\caption{Recommendation performance comparison of our model TSCR, TSCR w/o entity, and TSCR w/o item on TG-ReDial. `*' indicates significant improvements upon the best baseline in Fisher random test with $p$-value $<0.05$. Best performances are in bold. }
\label{tab:TSCR_seqablation2}
\centering
\begin{tabular}{lccc}
\toprule
\textbf{Model} & \textbf{Recall@1} & \textbf{Recall@10} & \textbf{Recall@50} \\
\midrule 
TSCR & \textbf{0.005*} & \textbf{0.032*} & \textbf{0.080*} \\%
\midrule
-w/o entity & 0.001 & 0.005 & 0.012 \\
-w/o item & 0.004 & 0.023 & 0.058 \\
\bottomrule
\end{tabular}
\end{table}
\subsection{Impact of Entity and Item in Sequence (RQ3)}
To understand what are the contributions of items and entities in a sequence, we conduct an ablation study comparing our model TSCR with its ablation variants (TSCR removing the non-item entities ``-w/o entity'' and TSCR removing the item mentions ``-w/o item''). 
The results on the ReDial dataset and TG-ReDial dataset are shown in Table~\ref{tab:TSCR_seqablation} and Table~\ref{tab:TSCR_seqablation2}, respectively. Observe from Table~\ref{tab:TSCR_seqablation} and Table~\ref{tab:TSCR_seqablation2}, both the item mentions and item-related entities in the conversation contribute to the final performance of TSCR. After removing item mentions or non-item entities from the context, the recommendation performance on all three metrics on the two datasets drops, which indicates the importance of the two components. Also, we observe that item mentions contribute more than non-item entities on the ReDial dataset. 
This might be because that item mentions are more reflective of user true preferences on the ReDial dataset and non-item entities contain more noise than item mentions. This suggests that sentiment analysis for entity mentions to distill the sequence of entity mentions might be beneficial~\citep{yan2021unified}. As for the TG-ReDial dataset, non-item entities are more useful than item mentions. One reason might be that the TG-ReDial dataset contains very few item mentions but much more non-item entities. The other reason is that the sequential dependency in item mentions in the TG-ReDial dataset is not as strong as in the ReDial dataset, given the TG-ReDial dataset contains items with multiple and diverse topics in a conversation. 

\begin{table}[t]
\caption{Recommendation performance comparison of our model TSCRKG, TSCRKG w/o entity, and TSCRKG w/o item on ReDial. `*' indicates significant improvements upon the best baseline in Fisher random test with $p$-value $<0.05$. Best performances are in bold. }
\label{tab:TSCRKG_seqablation}
\centering
\begin{tabular}{lccc}
\toprule
\textbf{Model} & \textbf{Recall@1} & \textbf{Recall@10} & \textbf{Recall@50} \\
\midrule 
TSCRKG & \textbf{0.087*} & \textbf{0.268*} & \textbf{0.465*} \\%
\midrule
-w/o entity & 0.081 & 0.255 & 0.431 \\%
-w/o item &  0.052 & 0.188 & 0.367 \\%
\bottomrule
\end{tabular}
\end{table}

\begin{table}[t]
\caption{Recommendation performance comparison of our model TSCRKG, TSCRKG w/o entity, and TSCRKG w/o item on TG-ReDial. `*' indicates significant improvements upon the best baseline in Fisher random test with $p$-value $<0.05$. Best performances are in bold. }
\label{tab:TSCRKG_seqablation2}
\centering
\begin{tabular}{lccc}
\toprule
\textbf{Model} & \textbf{Recall@1} & \textbf{Recall@10} & \textbf{Recall@50} \\%
\midrule 
TSCRKG & \textbf{0.007*} & \textbf{0.037*} & \textbf{0.090*}\\%
\midrule
-w/o entity & 0.004 & 0.016 & 0.045 \\
-w/o item & 0.006 & 0.030 & 0.073\\
\bottomrule
\end{tabular}
\end{table}

In addition, we conduct an ablation study comparing our model TSCRKG with its ablation variants (TSCRKG removing the non-item entities represented by ``-w/o entity'' and TSCRKG removing the item mentions represented by ``-w/o item''). The results on the ReDial dataset and TG-ReDial dataset are shown in Table~\ref{tab:TSCRKG_seqablation} and Table~\ref{tab:TSCRKG_seqablation2}, respectively. Observing from Table~\ref{tab:TSCRKG_seqablation} and Table~\ref{tab:TSCRKG_seqablation2}, we see a similar trend of TSCRKG with TSCR. That is, both the item mentions and non-item entities in TSCRKG are helpful because the performance on all evaluation metrics drops without any of them. Also, 
It is confirmed again that item mentions contribute more than non-item entities on the ReDial dataset while non-item entities contribute more than item mentions on the TG-ReDial dataset. 

\begin{figure}[t]
\centering
\includegraphics[width=0.7\columnwidth]{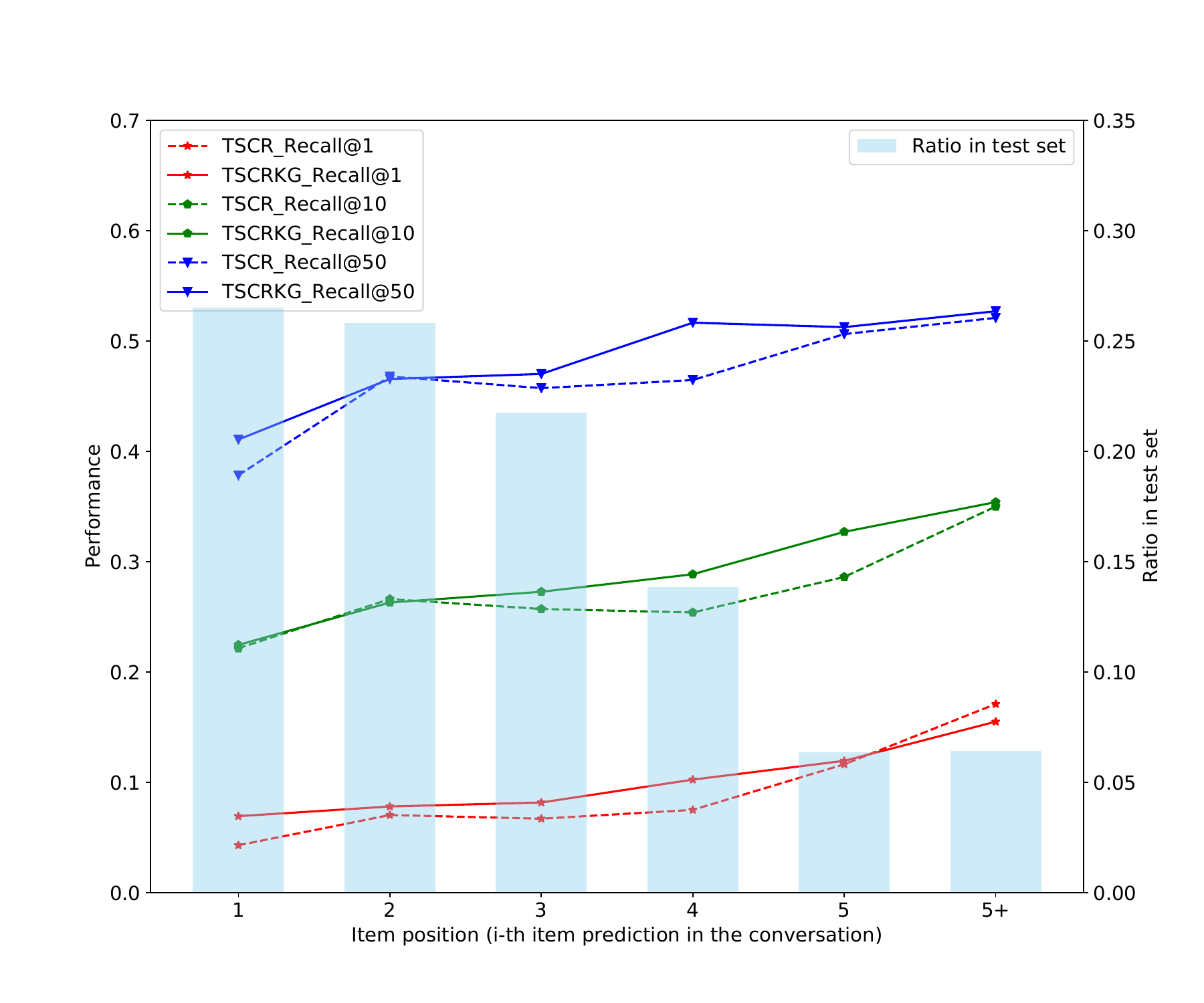}
\caption{The performance of TSCR and TSCRKG on ReDial with the ordinal number of item predictions.}
\label{fig:1}
\end{figure}

\begin{figure}[t]
\centering
\includegraphics[width=0.7\columnwidth]{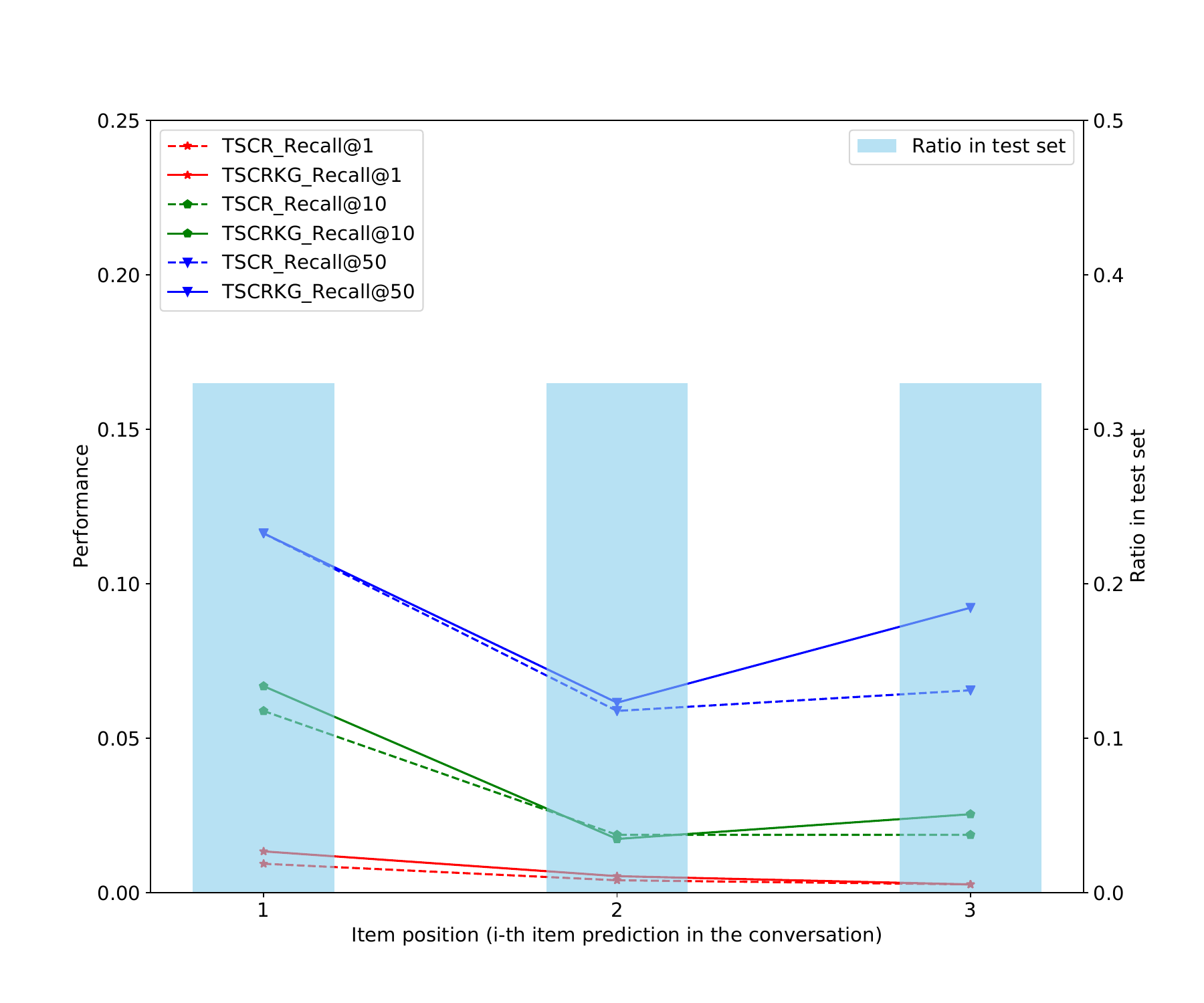}
\caption{The performance of TSCR and TSCRKG on TG-ReDial with the ordinal number of item predictions. }
\label{fig:2}
\end{figure}

\subsection{Effect of Item Position (RQ4)}
In this section, we explore whether the item position in the conversation affects the recommender performance. We first compare the recommender performance for each position of item predictions, from 1-st item prediction to 5+ item prediction in the conversation on the ReDial dataset. From Fig. \ref{fig:1}, we observe that, 
most of dialogs (74.1\%) on the ReDial dataset contain only 1--3 item recommendations. 
This is in line with that users expect the system can perform high-quality recommendations with fewer rounds in real applications. Overall, Fig. \ref{fig:1} shows that the performance of our TSCR model and TSCRKG improve as the item position increases. We attribute this to the fact that the models collect more contextual information about the user as the item position increases. A higher item position means a longer sequence length of item mentions and item-related entities. This indicates that both the performance of TSCR and TSCRKG improve when the sequence length gets longer, given the longer sequence in the conversation contains more useful information.  
Specifically, when the recommender suggests the first item (\ie the item position is equal to 1, corresponding to the classical problem
``cold start'' \citep{schein2002methods}), the TSCR and TSCRKG recommender can still achieve high performance based on the contextual information. In addition, TSCRKG outperforms TSCR in almost every item position, suggesting the robustness and effectiveness of TSCRKG.

Interestingly, compared to the performance of TSCR and TSCRKG for each position of item predictions from the ReDial dataset (Fig. \ref{fig:1}), we observe a very different trend from the TG-ReDial dataset, as shown in Fig. \ref{fig:2}. Unlike the ReDial dataset including a varying number of predicting items in each conversation, the TG-ReDial dataset contains three predicting items in each conversation. The performances of our TSCR model and TSCRKG drop from the 1-st item prediction to the 2-nd or 3-rd item prediction\footnote{To make sure the correctness of the observation, we check it by running one of the state-of-the-art baselines, $\text{C}^2$-CRS. We observe the same trend for $\text{C}^2$-CRS, i.e., the performance decreases from the 1-st item prediction to the 2-nd or 3-rd item prediction on the TG-ReDial dataset (e.g., 0.011, 0.001, and 0.001 on Recall@1 for the 1-st, 2-nd, and 3-rd item prediction, respectively.).}. This means that a longer sequence in the conversation of the TG-ReDial dataset contains more contextual information, yet, it does not always ensure better performance. This might be because that the TG-ReDial dataset in Chinese involves many more non-item entities than the ReDial dataset, leading to more noise. Also, given the TG-ReDial dataset contains simulated items with multiple and diverse topics in a conversation, the sequential dependency in item mentions in the TG-ReDial dataset is not as strong as in the ReDial dataset. Therefore, more item mentions in the user sequence do not guarantee better modeling of user true preferences. 

\begin{figure*}[t]
\centering
\begin{subfigure}[t]{0.47\textwidth}
\includegraphics[width=\columnwidth]{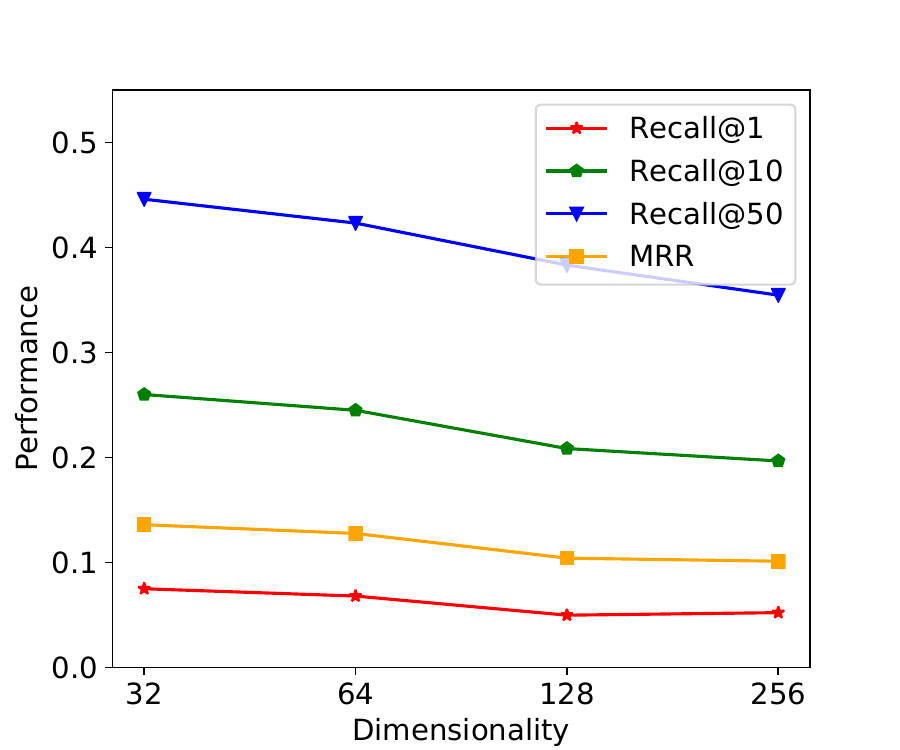}
\caption{Hidden dimensionality.}
\label{fig:dim}
\end{subfigure}
\begin{subfigure}[t]{0.47\textwidth}
\includegraphics[width=\columnwidth]{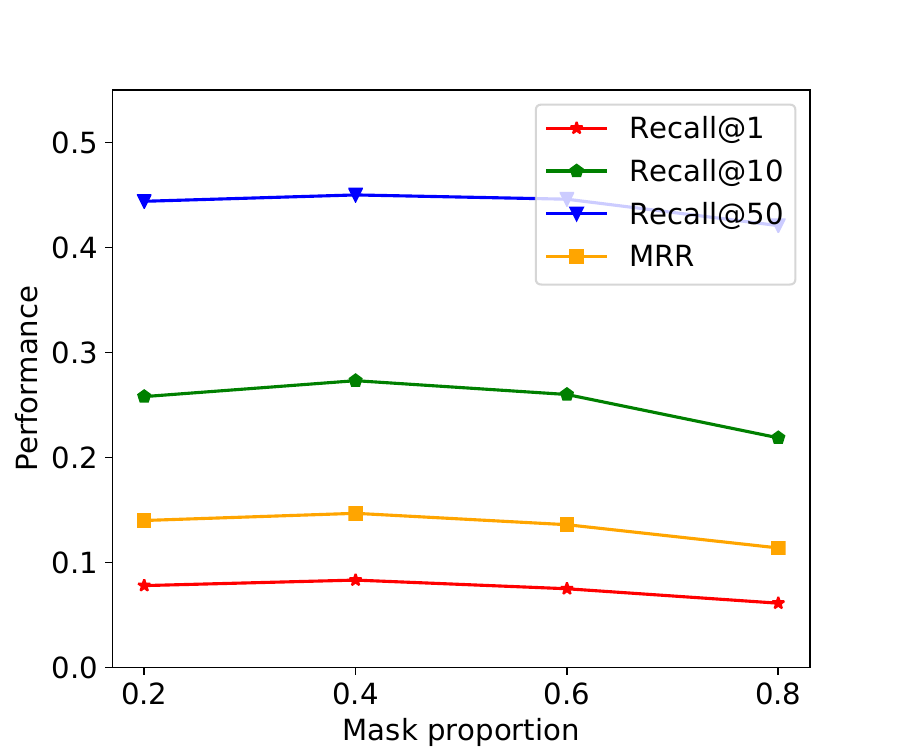}
\caption{Mask proportion.}
\label{fig:maskprop}
\end{subfigure}
\caption{Effect of hidden dimensionality and mask proportion.}
\end{figure*}

\subsection{Parameter Sensitivity (RQ5)}
In this section, we explore how our proposed model is affected by its main parameters, including the hidden dimensionality and the mask proportion. For simplicity, we use TSCR model and ReDial dataset for the experiments. 
\label{Para}
\paragraph{Effect of hidden dimensionality.}
We now explore how the hidden dimensionality affects the model performance. 
As shown in Fig. \ref{fig:dim}, we observe the recommendation performance of our TSCR model decreases with the embedding size increases. This is probably because of over-fitting. The TSCR model achieves the best performance when the hidden dimensionality is equal to 32. 
\paragraph{Effect of mask proportion.}
Fig. \ref{fig:maskprop} shows how the mask proportion affects the model performance. It can be seen that from Fig. \ref{fig:maskprop}, our TSCR model performs stably with the change of the mask proportion. 
The best-performing mask proportions are 0.4--0.6.  

\subsection{Case Study}
\label{sec:case}
\begin{table}[tb]
\caption{Cases extracted from the ReDial dataset.}
\label{tab:cases}
\begin{center}
\begin{tabular}{l|p{0.6\columnwidth}}
\toprule
\multicolumn{2}{c}{\textbf{Case1}}\\
\hline
\multirow{5}{*}{\textbf{Context}} & \textbf{Seeker}: Hey how are you?\\
& \textbf{Recommender}: Hi. I'm great.\\
& \textbf{Seeker}: Awesome! I'm looking for a good scary movie like \textit{\textbf{Rosemary's Baby (1968)}}, a classic, you know? \\
& \textbf{Seeker}: \textit{\textbf{The Exorcist (1973)}} is great too!\\
\hline
\multirow{2}{*}{\textbf{Ground truth}} & \textbf{Recommender}: Great movie. Have you seen \textit{\textbf{The Shining (1980)}} \\
\hline
\textbf{TSCR recommendation} & \textit{\textbf{The Shining (1980)}}\\
\hline
\textbf{TSCRKG recommendation} & \textit{\textbf{The Shining (1980)}}\\
\hline
\multirow{7}{*}{\textbf{Similar conversation}} & \dots\\
& \textbf{Seeker}: What about something like \textit{\textbf{The Exorcist (1973)}} or \textit{\textbf{Rosemary's Baby (1968)}}? I like classic horror a lot. \\
& \textbf{Recommender}: Definitely. Classic horror is good like \textit{\textbf{The Shining (1980)}}. \\
& \textbf{Seeker}: Loved that.\\
& \dots\\
\bottomrule
\end{tabular}
\begin{tabular}{l|p{0.6\columnwidth}}
\toprule
\multicolumn{2}{c}{\textbf{Case2}}\\
\hline
\multirow{5}{*}{\textbf{Context}} & \textbf{Recommender}: Hello, how are you? \\
& \textbf{Seeker}: Hi, I am good. \\
& \textbf{Recommender}: Looking for any good movie in particular? \\
& \textbf{Seeker}: I am looking for a nice comedy movie. \\
& \textbf{Seeker}: I love \textit{\textbf{Blades of Glory (2007)}}. \\
\hline
\multirow{2}{*}{\textbf{Ground truth}} & \textbf{Recommender}: Well, other good movies like that include \textit{\textbf{Talladega Nights: The Ballad of Ricky Bobby (2006)}}. \\
\hline
\textbf{TSCR recommendation} & \textit{\textbf{Wedding Crashers (2005)}}\\
\hline
\textbf{TSCRKG recommendation} & \textit{\textbf{Talladega Nights: The Ballad of Ricky Bobby (2006)}}\\
\hline
\textbf{Extracted knowledge graph} & \begin{minipage}{0.55\columnwidth}
\includegraphics[width=1\columnwidth]{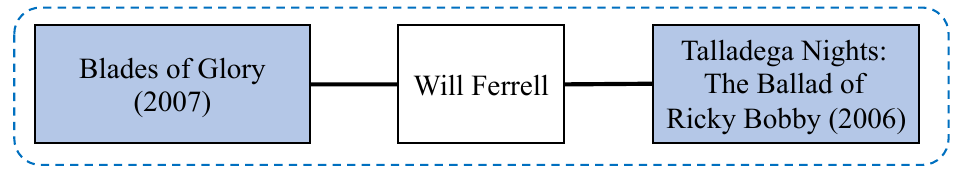}
\end{minipage}\\
\bottomrule
\end{tabular}
\end{center}
\end{table}

To illustrate the ability of TSCR and TSCRKG to generate reasonable recommendations, we present use cases of them on the ReDial dataset, as shown in Table \ref{tab:cases}. Observe from ``Case1'' in Table \ref{tab:cases}, both TSCR and TSCRKG can accurately recommend the movie, i.e., ``The Shining (1980)'', which is consistent with the ground truth. This is because both TSCR and TSCRKG are able to model the sequential dependency of contextual information. The two scary and classic movies mentioned in the context, i.e., ``Rosemary's Baby (1968)'' and ``The Exorcist (1973)'', are usually talked together and followed by another scary and classic movie ``The Shining (1980)''. This sequential dependency can be modeled and trained in similar conversations in the dataset. 

Observing from ``Case2'' in Table \ref{tab:cases}, we find that TSCRKG can generate the accurate recommendation with the ground truth, i.e., ``Talladega Nights: The Ballad of Ricky Bobby (2006)'', while TSCR is not able to generate the accurate recommendation. This is because modeling the sequential dependency is not sufficient for generating accurate recommendations in this case. There is no sequential dependency learned between the movies ``Blades of Glory (2007)''
and ``Talladega Nights: The Ballad of Ricky Bobby (2006)'', leading to a failure of TSCR. As a comparison, TSCRKG can generate the recommendation accurately, with the help of the knowledge graph. As shown in the row of ``extracted knowledge graph'' in Table \ref{tab:cases}, TSCRKG can learn from the knowledge graph by connecting the movies ``Blades of Glory (2007)'' and ``Talladega Nights: The Ballad of Ricky Bobby (2006)'' through the actor ``Will Ferrell''. This complete path in the knowledge graph leads to improved explainability for the recommended results. It confirms the effectiveness of TSCRKG and the benefit of incorporating the knowledge graph and sequential modeling. Furthermore, the contextual information learned from the knowledge graph can be used to enrich the conversation, i.e., the generated responses. For instance, one can enrich the response ``Well, other good movies like that include Talladega Nights: The Ballad of Ricky Bobby (2006).'' to the response ``Well, other good movies with Will Ferrel include Talladega Nights: The Ballad of Ricky Bobby (2006).'', by using the actor ``Will Ferrell'' from the knowledge graph. This suggests a potential research direction, i.e.,  enhancing the dialog generation model with extracted knowledge \cite{zheng2019enhancing}.

\section{Conclusion and Future Work}
\label{sec:conc}

In this paper, we have proposed the \acf{TSCR} for \acp{CRS}. TSCR deploys a Cloze task and models the sequential dependency of both items and entities in conversations by the deep bidirectional self-attention architecture. TSCR uses the knowledge base as a dictionary to get related entities, but does not use the structure of the knowledge base for any reasoning, making it simple and straightforward. Experimental results on the two CRS datasets show that our TSCR model, despite simple is highly effective, constituting a very strong baseline for future researchers to use. To take advantage of knowledge graphs, we have proposed an extension model TSCRKG, which fully uses the structure of knowledge graphs for offline representation learning and user sequence augmentation. The extensive experiments on the CRS datasets demonstrate the effectiveness of TSCRKG and the benefit of knowledge graphs. 

One limitation of this work is that we only focus on the recommender module. As for future work, we plan to incorporate the natural language response generation part as well. 
Moreover, in this work, we do not model the sentiment of mentioned items or entities and treat them as the same in the conversation. It is worth exploring the CRS by incorporating and modeling the sentiment (\eg positive or negative) from users' feedback in future work. Also, although we use entities from the DBpedia knowledge graph following previous \ac{CRS} work~\citep{chen2019towards,sarkar2020suggest,zhou2020improving,CRWalker}, the extracted entities may not be 100\% accurate. It is thus worth exploring the \ac{CRS} system by incorporating and modeling uncertainty and noise. 

In this work, we use the knowledge graph to initialize the offline representations of the augmented sequences, which are constructed by using the knowledge graph and conversation history. Then we learn the representations of the augmented sequences by bidirectional Transformer. In this way, the semantic representations from the knowledge graph are aligned with the conversation. However, one can also use other techniques, e.g., contrastive learning \cite{zhou2022c2}, to bridge the semantic gap between the knowledge graph and conversation history. Moreover, we model the sequential dependencies in all conversations. However, this is just the first step toward sequential modeling and there is still room for improvement. For example, conversations may involve complicated scenarios (e.g., multi-topics in case users shift focus from topic to topic) and do not form strict sequential dependencies in the entire sequence. A potential research direction is to model the entire sequence into several subsequences (e.g., detecting topic threads in multi-topic conversations \cite{zhou2020towards} and modeling each topic as a subsequence) so that more strict sequential dependencies remain in subsequences. It is also worth exploring the CRS model to detect sequential consistency in conversations, and thus the CRS model can predict the necessity to incorporate sequential modeling or model the noise in the sequence to relax the strict sequential dependencies. Last, exploring proactive CRS \citep{deng2024towards} and integrating other forms of conversational interactions are also promising directions.


\section{Acknowledgement}
This study is supported under the RIE2020 Industry Alignment Fund -- Industry Collaboration Projects (IAF-ICP) Funding Initiative, as well as cash and in-kind contribution from Singapore Telecommunications Limited (Singtel), through Singtel Cognitive and Artificial Intelligence Lab for Enterprises (SCALE@NTU). This research was also supported by the Natural Science Foundation of China (62402093), 
the NWO Smart Culture - Big Data / Digital Humanities (314-99-301), the NWO Innovational Research Incentives Scheme Vidi (016.Vidi.189.039), and the H2020-EU.3.4. - SOCIETAL CHALLENGES - Smart, Green And Integrated Transport (814961). 
All content represents the opinion of the authors, which is not necessarily shared or endorsed by their respective employers and/or sponsors. 

\clearpage
\bibliographystyle{ACM-Reference-Format}
\bibliography{bibfile}

\end{document}